# Uncertainties in modeling the capture process in heavy-ion collisions

I. I. Gontchar

*Omsk, Russia*

*e-mail: vigichar@hotmail.com*

In the present paper, we study the uncertainties in modeling the collision of complex nuclei ("heavy ions") resulting in capture of the nuclei into orbital motion. The effective interaction energy of the nuclei ("effective potential") consists of three terms: the Coulomb potential, the strong nucleus-nucleus potential, and the centrifugal term related to the orbital motion. The last term usually is considered in the literature as the simplest one. However, we found in the literature at least two different approaches for the centrifugal potential. To see the effect of using these different prescriptions, we evaluate the capture cross sections using the standard quantum-mechanical formula with the transmission coefficients calculated within the quasi-classical approximation. For the Coulomb- and strong nuclear terms we apply the semi-microscopical double-folding model with the effective nucleon-nucleon forces of Yukawa type. For the nucleon densities the two parameter Fermi formula is used with the parameters from the IAEA data base. Our calculations show that the two approaches for the centrifugal potential result in theoretical capture cross sections which are 20-40% different. This result holds for any collision energies.

Key words: fusion of atomic nuclei, double-folding model, effective potential, centrifugal term

## Introduction

Heavy-ion collisions resulting to the capture of interacting nuclei into orbital motion is the first step for formation of new superheavy elements and isotopes [1, 2]. Enormous number of works is devoted to the theoretical study of this step (see book [3], reviews [2, 4-8] and references therein). Typically, these works are focused on the bare strong nuclear part of the nucleus-nucleus interaction potential and/or on the couplings of the radial motion with the rotational and vibrational degrees of freedom of the colliding nuclides. After the reduction of the two-body problem to the one-body problem, the effective nucleus-nucleus interaction potential $U$ as the function of the distance between the centers of mass $R$ and of the system angular momentum $J$ reads

$$U(J,R) = U_C(R) + U_n(R) + U_{rot}(J,R). \qquad (1)$$

Here $U_C(R)$ denotes the Coulomb energy, $U_n(R)$ stands for bare strong nuclear interaction, and $U_{rot}(J,R)$ is the so-called rotational (centrifugal). This last term attracts our heed in the present work. Note, that typically the structure of $U_{rot}(J,R)$ is not discussed at all regarded to be trivial. Yet here we are focused on this term.

## 1. Review of the literature

After having analyzed significant amount of relevant works, we have found the following options. In many works [2, 9-23] the centrifugal energy reads:

$$U_{rot}(J,R) = \frac{\hbar^2 J(J+1)}{2m_R R^2}. \qquad (2)$$

In some works [7, 24-33] this term reads slightly different:

$$U_{rot\,p}(J,R) = \frac{\hbar^2 J^2}{2m_R R^2}. \qquad (3)$$

The subscript "p" in Eq. (3) denotes that colliding nuclei are considered as particles. We did not manage finding discussion of difference between Eqs. (2) and (3) in the literature. However, it is clear, that the difference is significant at small values of $J$, when the centrifugal term itself is insignificant. Accepting the version of Eq.(3), let us now focus on more important ambiguity.

Indeed, in some works [34 -37] the centrifugal energy is used in the following form:

$$U_{rot\ s}(J,R) = \frac{\hbar^2 J^2}{2(m_R R^2 + \mathcal{F}_P + \mathcal{F}_T)}. \tag{4}$$

the subscript "s" in Eq. (4) means that the system of two colliding nuclei is considered as a rigid (solid) body, i.e. rotation of each nucleus is synchronized with their orbital motion. $\mathcal{F}_P$, $\mathcal{F}_T$ stand for the moments of inertia of the projectile (P) and target (T) nuclei.

In almost all works employing Eq. (2) the collision process is considered in the quantum manner either by virtue of the single barrier penetration model or by the coupled channels approach. Thus, the appearance of $\hbar^2 J(J+1)$ for the absolute value of the orbital angular momentum (with $J$ to be the orbital quantum number) is natural. In these works, typically, possible rotation of the colliding nuclei is ignored. The works applying Eq. (3) the collision process is considered classically by means of the Hamilton equations accounting for dissipation and related thermal fluctuations. In such approach, colliding nuclei are expected to become gradually involved into rotation due to tangential friction. This circumstance is accounted for in [14, 32, 33]. In [7, 24] it was pointed out that the consideration is restricted by the nose-to-nose geometry. This implies rotation of the whole system as a rigid body and neglecting $\mathcal{F}_P$ and $\mathcal{F}_T$ may not be justified.

In Refs. [34, 36, 37] using Eq. (4) these moments of inertia are accounted for. However, the reason for considering rotation of the system as the rigid body at large values of $R$ is not clear.

## 2. Qualitative explanation of the uncertainty

The qualitative explanation of the uncertainty related to the use of the rigid body moment of inertia $\mathcal{F}_{sd} = m_R R^2 + \mathcal{F}_P + \mathcal{F}_T$ and of the moment of inertia of particles $\mathcal{F}_{pt} = m_R R^2$ seems to be as follows.

When the colliding nuclei are far away from each other their motion is purely translational. As the nuclei come closer their "surfaces" start interacting at the point of "touching". The interaction force possesses both radial and tangential terms. The latter makes nuclei rotating. The closer come nuclei the stronger the tangential term. Finally, the nuclei stick into rigid body («sticking limit»). Quantitatively this process is described by Eqs. (7-11) in Ref. [38] by Eqs. (7-11) in Ref. [39].

Thus, at large distances it is more correct using $\mathcal{F}_{pt}$, whereas at smaller distances it must be substituted by $\mathcal{F}_{sd}$. However, at the intermediate distances, a smooth transition from $U_{rot\ p}$ to $U_{rot\ s}$ should take place. In fact, in the intermediate case (im) the centrifugal term should read:

$$U_{rot\ im}(J,R) = \frac{\hbar^2 J^2}{2 m_R R^2} + \frac{\hbar^2 S_P^2}{2 \mathcal{F}_P} + + \frac{\hbar^2 S_T^2}{2 \mathcal{F}_T}. \tag{5}$$

Here two last terms correspond to the rotation of the projectile and target nuclei, whereas $S_P$ and $S_T$ denote their spins.

Let us shortly mention that as the nuclei approach each other, friction leads to melting the shell corrections which results in changing the superfluid expressions for the moment of inertia $\mathcal{F}_P$, $\mathcal{F}_T$ to the rigid body expressions [40]. In general, the problem of nucleus-nucleus collision of deformed nuclei accounting for all above mentioned peculiarities looks rather complex.

## 3. Model

We consider a simplified situation: collision of two spherical nuclei (projectile, P, and target, T) with charge numbers $Z_P$ and $Z_T$ and mass numbers $A_P$ and $A_T$. Interaction energy is evaluated trough Eq. (1) where $U_C$ and $U_n$ are calculated within the framework of the double-folding approach. The structure of both $U_C$ and $U_n$ is a double-folding integral involving two ingredients. The first one is the charge-, proton-, and neutron densities, $\rho_q(r), \rho_Z(r), \rho_N(r)$ where $r$ denotes the distance between the center of a nucleus and the interacting point (see Fig. 1).

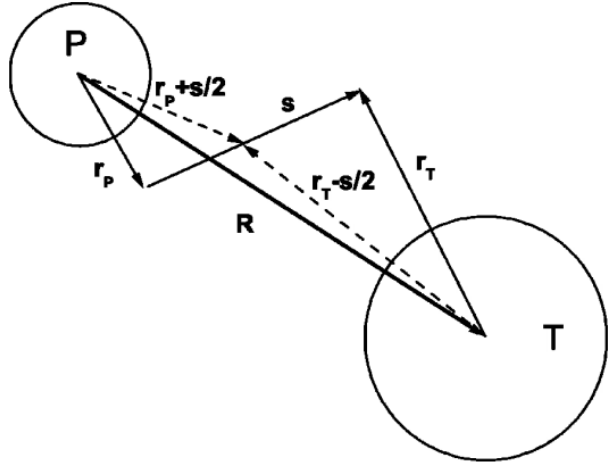

**Fig. 1**. Schematic illustration for Eq. (1).

The second ingredient is the two-point Coulomb and effective nuclear nucleon-nucleon forces (NN-forces). For the latter we use the Paris M3Y NN-forces [41] with the following direct part

$$v_D(s) = G_{D1}\left[\exp(-s/r_{v1})\right]/(s/r_{v1}) + G_{D2}\left[\exp(-s/r_{v2})\right]/(s/r_{v2}). \tag{6}$$

Here $G_{D1}$ =11062 MeV, $G_{D2}$ = −2537.5 MeV, $r_{v1}$ =0.25 fm, $r_{v2}$ =0.40 fm, $s$ denotes the distance between interacting points (see Fig. 1). For our purpose, the exchange part corresponding to pseudo potential with zero radius and modified strength can be successfully used [11]

$$v_{E\delta}(s) = G_{E\delta}\,\delta(\vec{s}). \tag{7}$$

Here $G_{E\delta} = -1040$ MeV fm$^3$ [11] instead of the original value −592 МэB fm$^3$ from the original work [41]. For details of more rigorous calculations with the finite range exchange forces see e.g. [42].

For the densities $\rho_N(r), \rho_Z(r)$, and $\rho_q(r)$ involved in the double-folding integrals we apply the two-parameter Fermi formula (2pF-formula):

$$\rho_F(r) = \frac{\rho_{CF}}{1 + \exp[(r - R_F)/a_F]}. \tag{8}$$

In Eq. (8) $R_F$ denotes the half-central density distance from the center of the nucleus; $a_F$ stands for the diffuseness of the density. The value of $\rho_{CF}$ is evaluated from the normalization condition. In the present work, parameters of the 2pF formula for protons (F=Z) and neutrons (F=N) are taken from [43]. As the nuclei come closer, the nucleon densities are supposed to be unchanged (the frozen densities approximation [10, 11, 29]). The moments of inertia $\mathcal{F}_P, \mathcal{F}_T$ are evaluated by integration in the same approximation and for the nucleon density the sum of proton and neutron densities is used.

For the charge density the radius parameter is taken to be equal to the proton radius, $R_q = R_Z$, whereas the charge diffuseness $a_q$ is calculated via $a_Z$ as follows (see [44]):

$$a_q = \sqrt{a_Z^2 + \frac{5}{7\pi^2}\left(0.76 - 0.11\frac{N}{Z}\right)}. \tag{9}$$

Thus, the charge density diffuseness surpasses the proton density one due to the finite charge distribution in proton and neutron.

Another source of information on the densities is the experimental review [45]. However, in that review only charge densities are presented and in addition the approximation (8) for the densities is often missing. Unfortunately, we did not find a more up-to-date review on this topic.

Fusion cross sections (CSs) $\sigma_{BPM}$ are evaluated according to the standard quantum-mechanical formula (see, e.g., [3]):

$$\sigma_{BPM} = \frac{\pi \hbar^2}{2 m_R E_{c.m.}} \sum_J (2J+1) T_J(E_{c.m.}). \tag{10}$$

Here $E_{c.m.}$ stands for the collision energy in the center-of-mass system; $J$ denotes the angular momentum in units of $\hbar$; $T_J(E_{c.m.})$ is the transmission coefficient. The latter is calculated either using the quasi-classical WKB-approximation

$$T_J(E_{c.m.}) = \left\{1 + \exp\left[\frac{2\Lambda_J(E_{c.m.})}{\hbar}\right]\right\}^{-1} \tag{11}$$

for the sub-barrier collision energies or within the parabolic barrier approximation

$$T_J(E_{c.m.}) = \left\{1 + \exp\left[\frac{2\pi(B_J - E_{c.m.})}{\hbar \omega_{BJ}}\right]\right\}^{-1} \tag{12}$$

for the above-barrier energies. In Eq. (11) $\Lambda_J(E_{c.m.})$ denotes the action calculated from the outer turning point up to the inner one. In Eq. (12) $B_J$ stands for the angular momentum dependent barrier height, $\omega_{BJ}$ is the corresponding "frequency".

## 4. Results

Let us start from considering what is the impact of the version of $U_{rot}(J,R)$ on the Coulomb barrier height $B_J$. For this aim we present in Fig. 2 three fractional differences

$$\xi_{Bp} = \frac{B_{Jp}}{B_0} - 1, \quad \xi_{Bs} = \frac{B_{Js}}{B_0} - 1, \quad \xi_{Bps} = \frac{B_{Jp}}{B_{Js}} - 1, \tag{13, 14, 15}$$

calculated for four reactions: $^{48}$Ca+$^{36}$S, $^{48}$Ca+$^{48}$Ca, $^{48}$Ca+$^{90}$Z, $^{48}$Ca+$^{124}$Sn. In Eqs. (13,14,15) $B_{Jp}$ denotes the barrier height calculated by means of Eq. (3) (particles version) at fixed value of $J$; $B_{Js}$ is the barrier height obtained using Eq. (4) (rigid body version). Figs. 2a,b show that the fractional rise of the barrier energy with $J$ is expressed much stronger for lighter systems. In Fig. 2c the impact of the option for calculating the rotational energy is demonstrated via the fractional difference $\xi_{Bps}$. The strongest effect is seen for the lightest system $^{48}$Ca+$^{36}$S.

Let us now see how the option for calculating the rotational energy influences the capture excitation functions. This is illustrated by Figs. 3,4,5,6 the design of which is the same. In panels a) and b) versus $E_{c.m.}/B_0$, the absolute values of the CSs are shown in the linear and logarithmic scales, respectively. The experimental CSs are shown by symbols, blue thin lines and red thick lines without symbols correspond to $\sigma_{BPMs}$ and $\sigma_{BPMp}$, respectively. In panels c) we display the fractions $\sigma_{BPMs}/\sigma_{BPMp}$ which demonstrate the effect of using the rigid body against particles version of $U_{rot}(J,R)$,

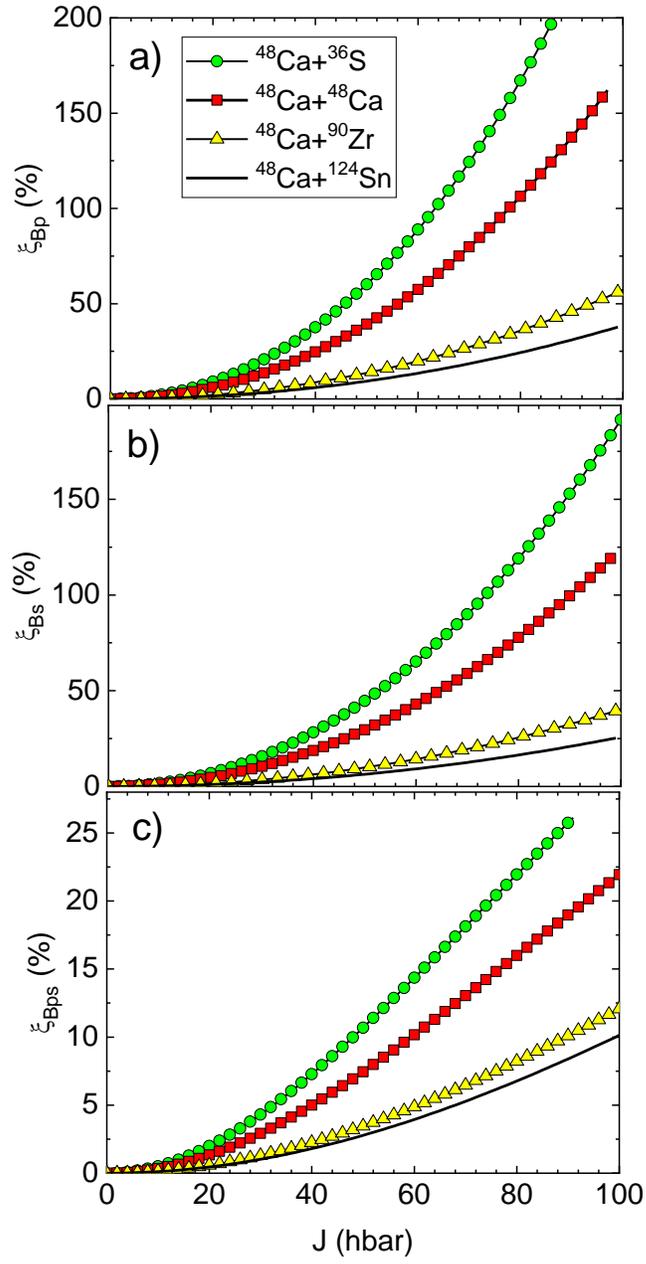

**Fig. 2.** Fractional differences of the barrier heights (see Eqs. (13,14,15)) versus the angular momentum. All notations are explained in panel a).

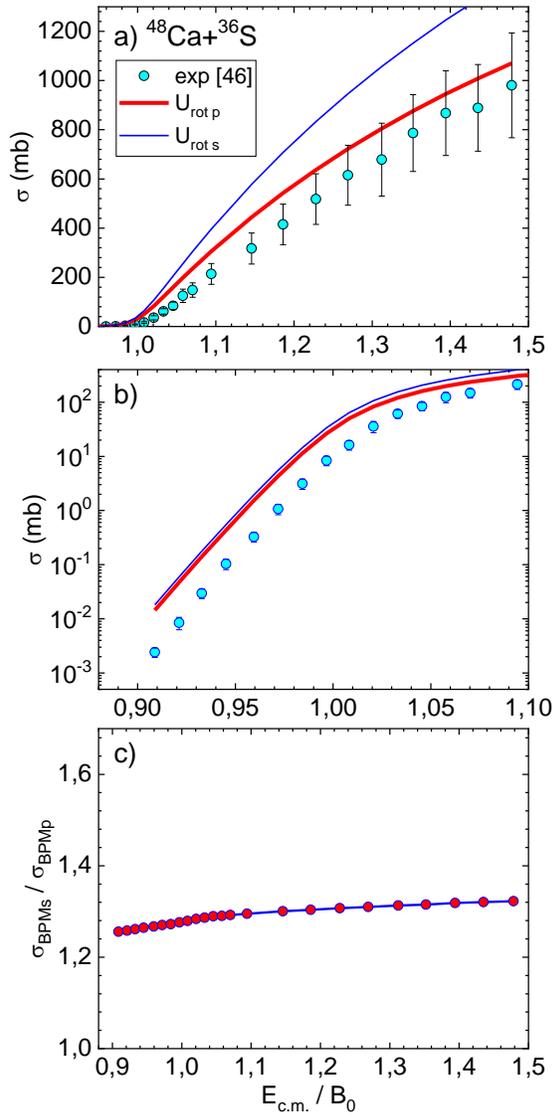

Fig. 3. For reaction $^{48}$Ca+$^{36}$S the following quantities are shown versus $E_{c.m.}/B_0$: a) CSs in the linear scale; b) CSs in the logarithmic scale; c) ratio $\sigma_{BPMs}/\sigma_{BPMp}$.

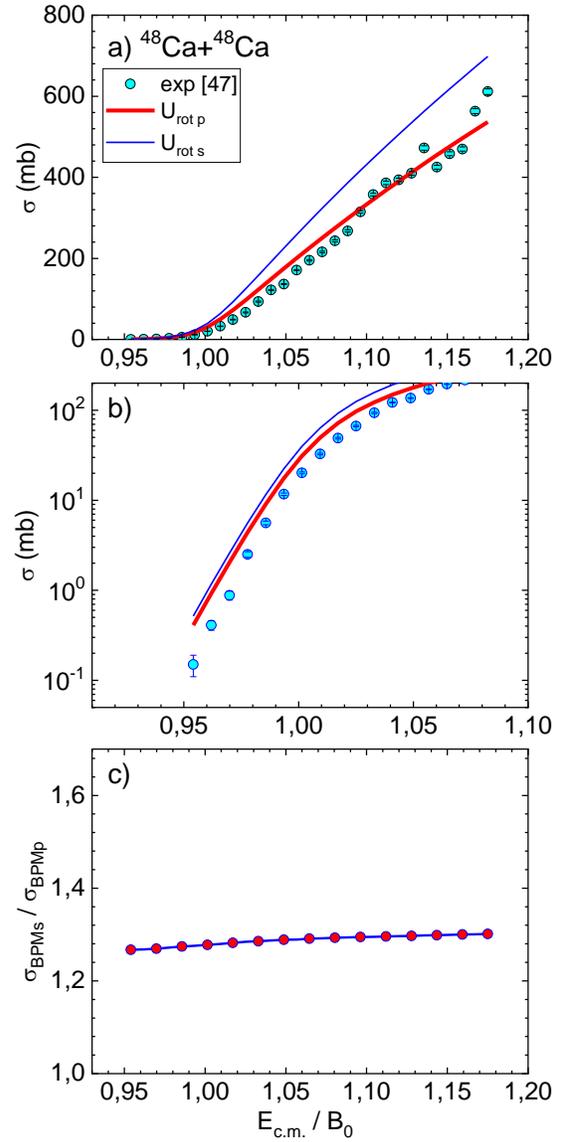

Fig 4. Same as in Fig. 3 but for reaction $^{48}$Ca+$^{48}$Ca. All notations as in Figs. 3.

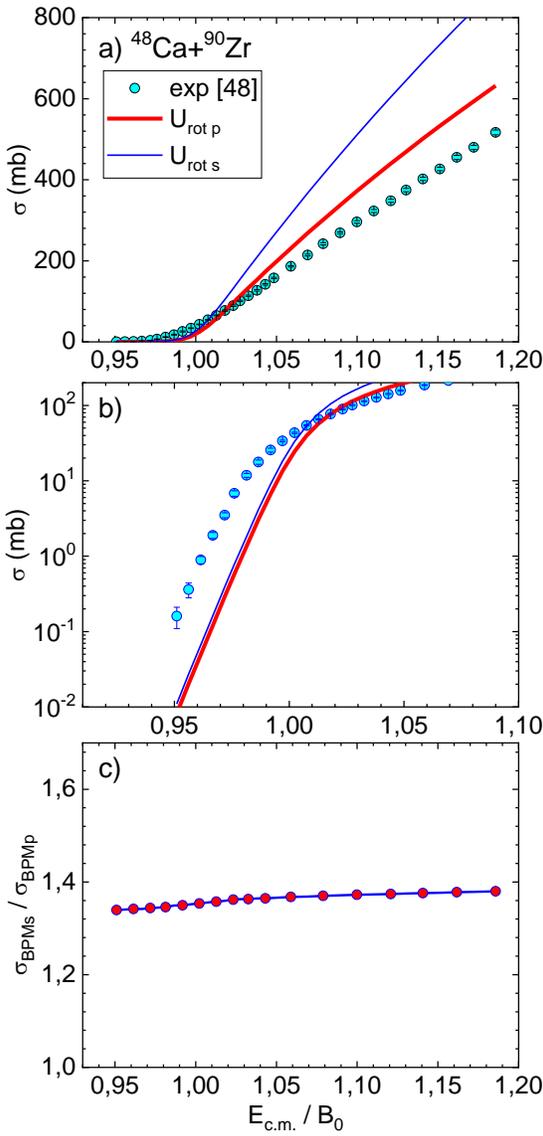
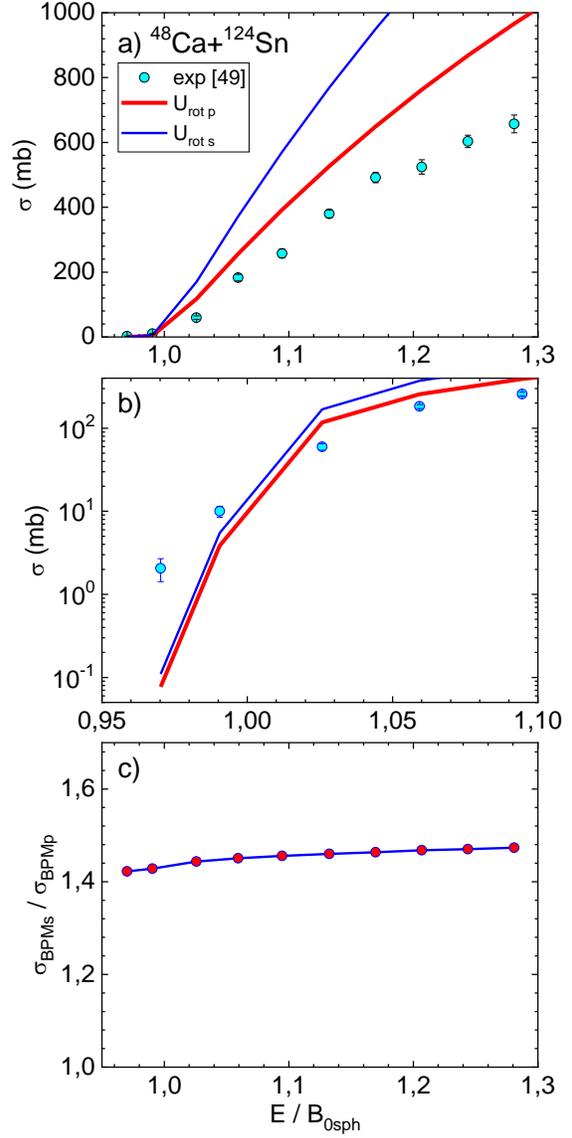

Fig 5. Same as in Figs. 3, 4 but for reaction $^{48}$Ca+$^{90}$Zr. All notations as in Figs. 3,4.

Fig 6. Same as in Figs. 3, 4, 5 but for reaction $^{48}$Ca+$^{124}$Sn. All notations as in Figs. 3,4,5.

As one could expect, for all reactions using Eq. (4) (rigid body) results in larger CSs in comparison to Eq. (3) (particles) just due to the lower barriers. But quantitatively this enhancement (i.e., the ratio $\sigma_{BPMs}/\sigma_{BPMp}$) is larger for the heaviest reaction $^{48}$Ca+$^{124}$Sn contrary to what could be expected from Fig. 2c.

To figure out this apparent contradiction we plot in Fig. 7 the values of the "critical" angular momenta $J_c$ for two options: Eq. (3) in panel a) and Eq. (4) in panel b). $J_c$ corresponds to the maximum in the partial capture CS versus $J$. One sees that, at the same values of $E_{c.m.}/B_0$, for heavier systems the critical angular momentum is significantly larger and overcompensates the lower values of $\xi_{Bps}$ in Fig. 2c.

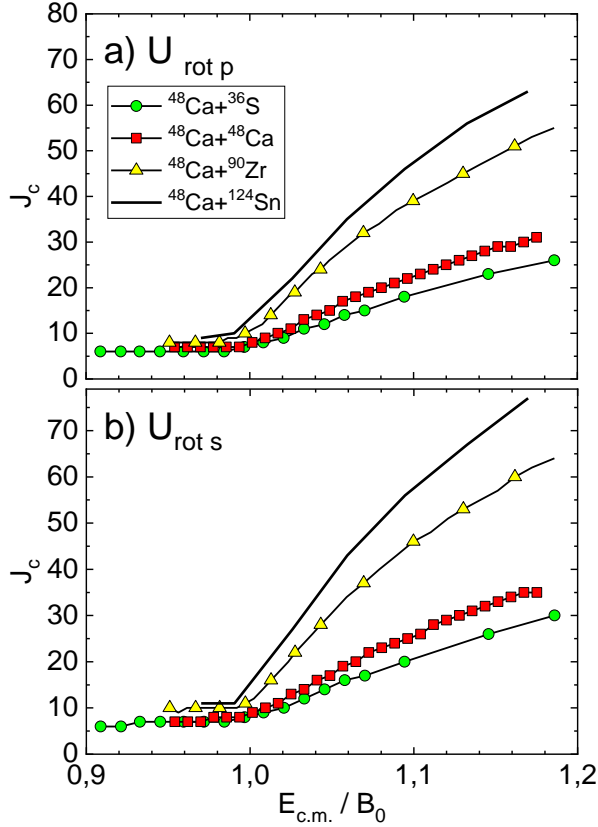

Fig. 7. Versus $E_{c.m.}/B_0$ the critical angular momenta shown versus $J_c$ is shown for two options: Eq. (3) in panel a) and Eq. (4) in panel b).

## 4. Conclusions

There are at least two significantly different recipes in the literature for calculating the so-called rotational or centrifugal part of the effective potential energy in heavy-ion collisions. The first one given by Eq. (3) (particles) seems to ignore the finite size (radius) of the colliding nuclei whereas in the second one given by Eq. (4) (rigid body) this effect is accounted for. These two versions seem to represent two limiting cases corresponding to significantly different center-of-masses distances.

To clarify quantitatively to what extent this uncertainty is significant, we have calculated the capture cross sections for four reactions: $^{48}Ca+^{36}S$, $^{48}Ca+^{48}Ca$, $^{48}Ca+^{90}Zr$, $^{48}Ca+^{124}Sn$. These calculations have been performed by means of the standard quantum-mechanical formula (see Eq. (10)) with the transmission coefficients evaluated within the WKB-approximation. The Coulomb and strong nuclear terms of the nucleus-nucleus effective potential the double-folding model with the Paris M3Y effective nucleon-nucleon forces has been applied. For the nucleon densities we have used the two-parameter Fermi formula with the parameters from the IAEA data base.

The calculations have demonstrated that the particles version results in higher barriers and smaller values of critical angular momentum. The second effect turns out to be stronger and thus the theoretical CSs within the rigid body version are 20-40% larger comparing to the particles version. This result does not depend upon the collision energy.

The author is indebted to Dr. M.V. Chushnyakova and Dr. V.L. Litnevsky for fruitful discussions.